\documentstyle[twocolumn,aps,prl,epsf,floats]{revtex}

\begin{document}

\draft

\twocolumn[\hsize\textwidth\columnwidth\hsize\csname@twocolumnfalse\endcsname

\title{Effective Area-Elasticity and Tension of Micro-manipulated Membranes}

\author{J.-B. Fournier$^1$, A. Ajdari$^1$ and L. Peliti$^2$}
\address{$^1$ Laboratoire de Physico-Chimie Th\'eorique, CNRS Esa~7083, ESPCI,
10 rue Vauquelin, F-75231 Paris C\'edex 05, France}
\address{$^2$ Dipartimento di Scienze Fisiche and Unit\`a INFM, Universit\`a
``Federico II'', Complesso M. S. Angelo, I--80126 Napoli (Italy)}
\date{\today}
\maketitle
\begin{abstract}
We evaluate the effective Hamiltonian governing, at the optically resolved
scale, the elastic properties of micro-manipulated membranes. We identify
floppy, entropic-tense and stretched-tense regimes, representing different
behaviors of the effective area-elasticity of the membrane. The corresponding
effective tension depends on the microscopic parameters (total area, bending
rigidity) and on the optically visible area, which is controlled by the imposed
external constraints.  We successfully compare our predictions with recent data
on micropipette experiments.  
\end{abstract}
 \vspace*{-.3cm}
\twocolumn]\narrowtext


Micro-manipulation techniques, where, e.g., magnetic or optical traps pull on
micrometric beads attached to a material, are increasingly used to probe the
elastic response of various soft-matter systems, such as biological polymers
(e.g., DNA~\cite{Strick96} or proteins~\cite{Kellermayer97}), phase boundaries
of Langmuir monolayers~\cite{Wurlitzer00}, surfactant vesicles~\cite{Helfer00}
and even living cells~\cite{Hochmuth00}. Combined with optical observations,
these techniques should allow in the future to finely test the elastic theories
of complex systems, e.g, by monitoring the shape and fluctuations of a system
while imposing inhomogeneous boundary conditions. Although surfactant
membranes and vesicles~\cite{book,Seifert_revue} are apparently one of the
simplest system, there is still some confusion regarding the appropriate
Hamiltonian describing their elasticity, and in particular about the role and
value of an {\em effective tension\/}~$\sigma$. The latter is not a
well-identified microscopic quantity, contrary to the bending rigidity~$\kappa$.
In different thermodynamic ensembles~\cite{David91}, various approaches have
been proposed: phenomenological self-consistent
theories~\cite{Helfrich84,Evans90,Seifert95}, approximations involving Lagrange
multipliers~\cite{Seifert95,Milner87}, or formal renormalization
schemes~\cite{David91,Peliti85,Kleinert86,Cai95,Cai94}.  In this Letter, we
propose an approach based on a large but finite
coarse-graining~\cite{Lubenskybook,JBF99}, devised for the interpretation of
measurements combining micro-manipulation and optical observations.

Fluid membranes in aqueous solutions often consist of a fixed number of
highly insoluble lipid or surfactant molecules. Since stretching a flat
membrane involves very high energies, while macroscopically bending it
involves energies of order $k_{\rm B}T$, membranes are commonly modeled
as fluctuating two-dimensional sheets with a prescribed microscopic
area~$\bar A$ and a curvature elasticity~\cite{Seifert_revue}. At the
{\em macroscopic scale\/}, membranes actually appear very different:
their optically visible area~$A$ fluctuates about some value depending
on the temperature and on the external constraints. Part of the total
area $\bar A$ is stored in short scale fluctuations that are optically
unresolved~\cite{Helfrich84,Evans90}.  For such a critically fluctuating
system, a coarse-grained {\em effective\/} Hamiltonian ${\cal H}_{\rm
eff}$, integrating all sub-optical details, is clearly more adequate
than the microscopic Hamiltonian.

Our goal is to calculate this effective macroscopic Hamiltonian ${\cal
H}_{\rm eff}$ and to investigate the associated area-elasticity and
tension. We start by considering a quasi-planar membrane with a fixed
microscopic area~$\bar A$, which is attached to a fixed frame of
area~$L^2$.  We choose the simplest microscopic curvature Hamiltonian:
the lowest-order, quadratic approximation of the Canham-Helfrich
Hamiltonian~\cite{Canham70,Helfrich73}:
\begin{equation}
{\cal H}_c[h_m]=\int\!d^2x\,\frac{\kappa}{2}\left(\nabla^2 h_m\right)^2.
\end{equation}
Here $h_m({\bf x})$ is the height of the membrane above a reference plane (Monge
gauge), as resolved microscopically.  We then determine the coarse-grained
Hamiltonian ${\cal H}_{\rm eff}[h]$, where $h({\bf x})$ is the height of the
membrane as resolved optically. This Hamiltonian is such that $\exp(-\beta{\cal
H}_{\rm eff}[h])$ gives the probability for the occurrence of any optically
visible membrane shape $h$, whatever its fluctuating microscopic detail.
Technically, this is a one-step renormalization of the fixed area constraint. 

We find that ${\cal H}_{\rm eff}$ involves a {\em non-linear\/}
area-elasticity energy ${\cal H}_s(A)$ for the coarse-grained, optically
visible, area $A$. This is the effective potential which is probed by
pulling a membrane in an optically resolved micro-manipulation.
Depending on the microscopic excess area $\alpha_m\!=\!(\bar A-L^2)/L^2$
and on the constraints exerted on the membrane, we find three distinct
regimes: a floppy regime, an entropic-tense regime, and a
stretched-tense regime. We provide explicit formulae for the effective
tension $\sigma(A)\equiv d{\cal H}_s/dA$ in these three regimes.  To
better describe the tense regime, we further incorporate a microscopic
stretching elasticity.  We then contrast our approach and the resulting
picture with the common use of a heuristic tension proposed by Helfrich
and Servuss~\cite{Helfrich84}.  Eventually, we point out that our
results can be applied to giant vesicles with a fixed
volume~$V_0=\frac{4}{3}\pi R_0^3$, by taking $L^2=4\pi R_0^2$.
This allows us to perform a first test of our theory: re-analyzing the
micro-pipette experiments of Evans and Rawicz~\cite{Evans90,Rawicz00},
we find an excellent fit for the cross-over between the entropic-tense
and the stretched-tense regimes. Finally, we point out the
approximations involved in our calculations and we propose several
possible experiments. 

\begin{figure}
\centerline{\epsfxsize=6cm\epsfbox{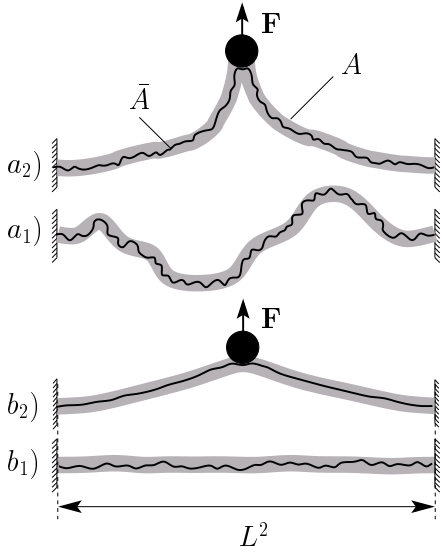}}
\caption{Membranes with a microscopic area~$\bar A$ and an optically
visible area~$A$, attached to a fixed frame of area~$L^2$. The
effective, optically measurable tension of an initially ``flat''
membrane~($b_1$), having a small microscopic excess area $\alpha_m=(\bar
A-L^2)/L^2$, can be further increased by an external
perturbation~($b_2$).  Similarly, an initially ``floppy''
membrane~($a_1$) can be externally led into a tense regime~($a_2$).}
\label{fig:membranes}
\end{figure}

The thermal fluctuations of a membrane of fixed microscopic area $\bar A \simeq
\int\!d^2x\,(1+\frac{1}{2}\left(\nabla h_m\right)^2)$ are described by the
partition function
\begin{equation}\label{Z}
Z=\int\!{\cal D}[h_m]\,\delta\!\left(L^2\!+\!\int\!d^2x\,
\frac{1}{2}\left(\nabla
h_m\right)^2 - \bar A\right)e^{-\beta{\cal H}_c}.
\end{equation}
We perform the decomposition $h_m({\bf x})=h({\bf x})+h^>({\bf x})$, in which
$h({\bf x})$ (in gray in Fig.~\ref{fig:membranes}) has wavevectors in the range
$0<q<\Lambda$, where $\Lambda$ corresponds to the experimental limit of
resolution, e.g., optical, and $h^>({\bf x})$ has wavevectors in the range
$\Lambda<q<\Lambda_0$, where $\Lambda_0$ is a molecular cutoff. Splitting the
integration as ${\cal D}[h_m]={\cal D}[h]{\cal D}[h^>]$, and writing the
delta-function in Fourier space, we can rewrite Eq.~(\ref{Z}) as
\begin{equation}
Z=\int\!{\cal D}[h]\exp\left\{-\beta\left[
{\cal H}_c[h]+{\cal H}_s(A)
\right]\right\}
\end{equation}
where $A[h]=L^2\!+\!\int\!d^2x\,\frac{1}{2}\left(\nabla
h\right)^2$ is the optically visible area, and
\begin{eqnarray}\label{precol}
{\cal H}_s(A)=-\frac{1}{\beta}\ln\!
\int_{-i\infty}^{i\infty}\!d\lambda&&\,
\exp\left[
-\lambda\left(\bar A-A\right)\right.\nonumber\\
&&\left.-\frac{L^2}{2}\int_{\bf q^>}\!\!
\ln\left(\beta\kappa q^4-\lambda q^2\right)
\right].
\end{eqnarray}
This defines the effective Hamiltonian at the optical scale ${\cal H}_{\rm
eff}[h]={\cal H}_c[h]+{\cal H}_s(A[h])$. In the thermodynamic limit, the above
integral can be evaluated at the saddle point:
\begin{equation}
{\cal H}_s(A)\simeq\frac{\lambda_s}{\beta}\left(\bar A-A\right)
+\frac{L^2}{2\beta}\int_{\bf q^>}\!\!
\ln\left(\beta\kappa q^4-\lambda_s q^2\right),
\end{equation}
with $\lambda_s$ the solution of $\bar A-A=\frac{1}{2}L^2\int_{\bf q^>}\!
(\beta\kappa q^2-\lambda_s)^{-1}$.  The effective tension for an optical area
$A$, $\sigma(A)\equiv d{\cal H}_s/dA=-\lambda_s/\beta$, is thus related to the
area stored in the sub-optical modes by: 
\begin{equation}\label{sp1}
\frac{\bar A-A}{L^2}
\simeq
\frac{1}{8\pi\beta\kappa}\ln
\frac{\kappa\Lambda_0^2+\sigma}{\kappa\Lambda^2+\sigma}.
\end{equation}
Integrating, we obtain an explicit formula for the surface potential
(see Fig.~\ref{fig:hs2a}):
\begin{equation}\label{eq:hs2a}
{\cal H}_s(A)=\frac{L^2\Lambda_0^2}{8\pi\beta}
\left[\left(\frac{\Lambda^2}{\Lambda_0^2}-1\right)\ln\left(e^{X}-1\right)
+X\right],
\end{equation}
where $X=8\pi\beta\kappa(\bar A-A)/L^2$. ${\cal H}_s(A)$
has a minimum for
\begin{equation}
A^\star=\bar A-\frac{L^2}{4\pi\beta\kappa}\ln\frac{\Lambda_0}{\Lambda},
\end{equation}
at which the tension $\sigma$ vanishes.  For $A<A^\star$, the tension saturates
to the {\em negative\/} value $\sigma\simeq-\kappa\Lambda^2$, while for
$A\to\bar A^-$ it diverges as $\sigma\simeq(\Lambda_0^2L^2)/[8\pi\beta(\bar
A-A)]$, as a consequence of the prescribed microscopic area~$\bar A$.

\begin{figure}
\centerline{\epsfxsize=8cm\epsfbox{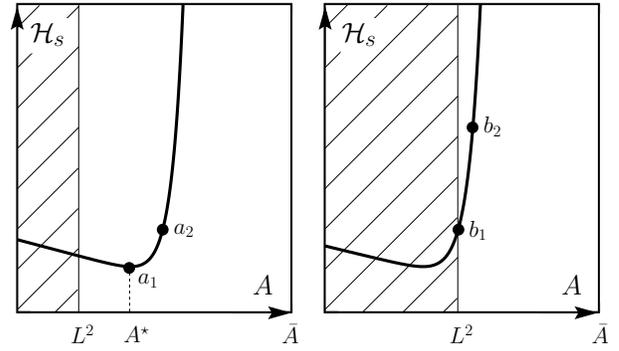}}
  \caption{Coarse-grained area-elasticity ${\cal H}_s(A)$ as a function of the
apparent area $A$. Since $A>L^2$, the hatched region is physically unaccessible.
Depending on the microscopic excess area, the membrane is initially floppy
($a_1$) or tense ($b_1$). Further stretching can be externally induced
($a_2,b_2$).  The floppy membrane ($a_1$) can, e.g., be led into a state of
tension ($a_2$) similar to that of the unperturbed tense one ($b_1$).}  
\label{fig:hs2a}
\end{figure}


\medskip{\em Floppy and tense regimes\/}. -- A crucial remark is that
the optical area $A$ cannot be smaller than the frame area $L^2$, which
results in two possibilities (Fig.~\ref{fig:hs2a}).  If the microscopic
excess area $\alpha_m=(\bar A -L^2)/L^2$ is larger than
\begin{equation}
\alpha_m^c=\frac{1}{4\pi\beta\kappa}\ln\frac{\Lambda_0}{\Lambda},
\end{equation}
we have $A^\star>L^2$ and the minimum of ${\cal H}_s(A)$ is indeed
physically realizable [Fig.~2\,$(a_1)$]. The coarse-grained unperturbed
membrane is then in a {\em floppy\/} state [Fig.~1\,$(a_1)$], in
which the tension has a vanishing optimal value. 

Conversely, if the attachment to the frame results in
$\alpha_m\!<\!\alpha_m^c$, the unphysical hatched region of
Fig.~\ref{fig:hs2a} comes to the right of the minimum, and the
zero-tension state is no longer accessible. The unperturbed
coarse-grained membrane finds then its minimum energy in a {\em
tense\/}, optically flat state, with $A=L^2$ [see Figs.~2\,$(b_1)$ and
1\,$(b_1)$].
Two tense regimes are however possible: (i)~when
$\kappa\Lambda^2\ll\sigma\ll \kappa\Lambda_0^2$, the membrane is in an
{\em entropic-tense\/} state with (still in the flat state):
\begin{equation}\label{sigma0}
\sigma_0\simeq\kappa\Lambda_0^2\,\exp\left(-8\pi\beta\kappa\alpha_m\right),
\end{equation} 
according to Eq.~(\ref{sp1}). While $\kappa\Lambda_0^2$ compares with ordinary
liquids tensions, the exponential reduction factor in Eq.~(\ref{sigma0}) leads
to extraordinary small values of membrane tensions, in line with many
observations~\cite{Seifert_revue,Faucon89}. (ii)~When $\sigma$ compares with
$\kappa\Lambda_0^2$, due to the smallness of $\alpha_m$, the membrane is in a
{\em stretched-tense} regime, where the tension describes the divergence of
${\cal H}_s$ for $A\rightarrow\bar A^-$, i.e.,
$\sigma\simeq\Lambda_0^2/(8\pi\beta\alpha_m)$. (This behavior will be corrected
next by the introduction of a microscopic streching elasticity.) A remarkable
point is that in the tense states, $\sigma$ is independent of the
coarse-graining scale, $\Lambda$.  Besides, since $\alpha_m<\alpha_m^c$, the
characteristic length $(\kappa/\sigma)^{1/2}$ is always smaller than the optical
cutoff, $\Lambda^{-1}$, thus the bending rigidity is masked by the effective
tension.  

If now the membrane is {\em stretched by external means\/}, the typical value of
$A$ is increased compared to its free value ($A^\star$ or $L^2$), and the system
can be brought into tenser regimes.  The response to weak perturbations of an
initially floppy membrane [$\alpha_m>\alpha_m^c$, Figs.~1\,$(a_1)$ and
2\,$(a_1)$] is described, from Eq.~(\ref{eq:hs2a}), by a quadratic elasticity
around the minimum $A^\star$:
\begin{equation}
{\cal H}_s(A)\simeq\frac{1}{2}k_{\rm eff}L^2
\left(\frac{A-A^\star}{L^2}\right)^2,\quad
k_{\rm eff}=8\pi\beta\kappa^2\Lambda^2,
\end{equation}
Under stronger stretching, the membrane reaches an {\em entropic-tense} state
[Figs.~1$(a_2)$ and~2$(a_2)$]. Its effective elasticity is then similar to that
of an initially tense membrane. According to Eq.~(\ref{sp1}), the tension is
given by
\begin{equation}\label{sigmaexp}
\sigma\simeq\sigma_0\exp\left(8\pi\beta\kappa\alpha\right),
\end{equation}
where $\alpha\!=\!(A-L^2)/L^2$ is the {\em apparent\/} excess area. Eventually,
further pulling brings the membrane into the stretched-tense regime with
$\sigma\simeq\Lambda_0^2/[8\pi\beta(\alpha_m-\alpha)]$.


\medskip{\em Including Stretching Elasticity\/}. -- 
A better description of a
strongly stretched membrane can actually be achieved by taking into account the
small but finite extensibility at microscopic scales.  We remove the
delta-function in Eq.~(\ref{Z}), and replace the Hamiltonian by
\begin{equation}
{\cal H}_c+
\frac{k_m}{2\bar A}\left[
L^2\!+\!\int\!d^2x\,\frac{1}{2}\left(\nabla h_m\right)^2-\bar A
\right]^2.
\end{equation}
Applying a Hubbard-Stratonovich transformation~\cite{Lubenskybook}, we obtain
Eq.~(\ref{precol}) with an additive correction $\frac{1}{2}(\bar A/\beta
k_m)\lambda^2$ in the exponential. Thus, instead of Eq.~(\ref{sp1}), the
saddle-point equation becomes (with $\sigma=-\lambda_s/\beta$):
\begin{equation}\label{sp2}
\bar A\left(1+\frac{\sigma}{k_m}\right)-A=
\frac{L^2}{8\pi\beta\kappa}\ln\frac{\kappa\Lambda_0^2+\sigma}
{\kappa\Lambda^2+\sigma}.
\end{equation}
Integrating this equation leads to an improved form for ${\cal H}_s(A)$.
Clearly, as long as $\sigma\ll k_m$, Eqs.~(\ref{sp1}) and~(\ref{sp2}) are
equivalent. The present correction is useful only to describe the
cross-over to and the tense-stretched regime [$\sigma$ comparable to or
larger than $\mathrm{min}(k_m,\kappa\Lambda_0^2)$], when it modifies the
divergence of ${\cal H}_s$ for large values of $A$.


\medskip{\em Comparison with the model of Helfrich and Servuss\/}.~-- In
Ref.~\cite{Helfrich84}, the membrane is macroscopically depicted as a
flat surface ($A\equiv L^2$), and $\sigma$ is introduced as a {\em
microscopic\/} surface tension, which is self-consistently determined by
prescribing the average value of the fluctuating microscopic excess-area
(in the ensemble in which the frame is fixed).  Then the contribution of
the stretching elasticity is added by hand. In our theory, we take into
account the actual microscopic membrane elasticity, and we define the
effective tension $\sigma(A)$ as the derivative of the area elasticity
${\cal H}_s(A)$ associated with the {\em coarse-grained\/} membrane
area~$A$.  Although Eqs.~(\ref{sp1}) and~(\ref{sp2}) are very similar to
those derived in Ref.~\cite{Helfrich84}, our clear distinction between
$L^2$, $A$ and $\bar A$, naturally allows to describe the area
elasticity of a membrane deformed by external actions (while the
description of Ref.~\cite{Helfrich84} allows no distinction between $A$
and $L^2$).  

\medskip{\em Quasi-spherical vesicles\/}. -- Let us now briefly discuss
the case of a quasi-spherical vesicle with a fixed volume
$V_0=\frac{4}{3}\pi R_0^3$.  Parameterizing its shape by
$R(\theta,\phi)=R_0\,[1+\sum_{\ell m}u_{\ell m}Y_{\ell m}(\theta,\phi)]$
and taking into account the volume constraint
$\sqrt{4\pi}\,u_{0,0}\!=\!-\sum_{\ell>0,m}|u_{\ell m}|^2$, the
microscopic area is fixed, in the partition function, by the factor
(see, e.g.,~\cite{Seifert95}):
\begin{equation} 
\delta\!\left(
4\pi R_0^2+R_0^2\!\sum_{\ell\ge2,m}\!
\frac{1}{2}(\ell-1)(\ell+2)|u_{\ell m}|^2-\bar A
\right).
\end{equation}
Comparing with the delta-function in Eq.~(\ref{Z}), we see that the area $4\pi
R_0^2$ plays the role of the frame area $L^2$. It will be shown in detail
elsewhere that a coarse-graining procedure yields the same results as in the
flat case, provided the vesicle is large enough for the optical wavelength to
corresponds to quasi-planar modes.  At the optical resolution, the vesicle still
has a prescribed volume $\simeq\!V_0$ (the volume is almost unaffected by
quasi-planar modes), and is described by the effective Hamiltonian ${\cal
H}_{\rm eff}$ obtained previously, upon replacement of $L^2$ by $4\pi R_0^2$.

\begin{figure}
\centerline{\epsfxsize=7cm\epsfbox{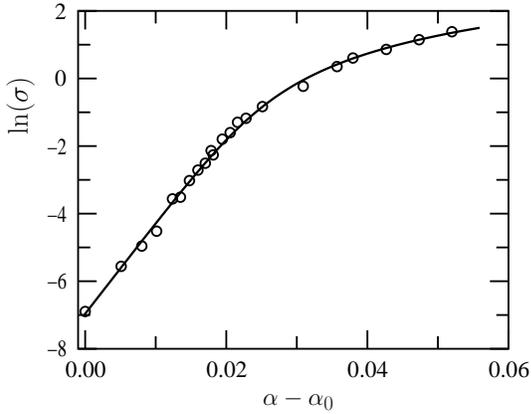}}
\caption{Fit by Eq.~(\ref{forfit}) of the data obtained in a recent
micropipette experiment by the group of Evans~\protect\cite{Rawicz00}
(see text).} 
\label{fig:fit}
\end{figure}

To check our model, we have re-analyzed the most recent micropipette
experiments~\cite{Rawicz00}: a certain amount of membrane area, initially
invisible at the optical resolution, is aspirated from a vesicle my means of a
pressurized micropipette. The induced tension~$\sigma$, obtained from Laplace's
law, is measured as a function of the variation of the apparent area.
Our Eq.~(\ref{sp2}) can be rewritten in the tense
regime as~\cite{noteeq16}
\begin{equation}\label{forfit}
\alpha=C_1+\left(8\pi\beta\kappa\right)^{-1}\ln\sigma+C_3\frac{\sigma}{k_m},
\end{equation}
with $C_1=\alpha_m-(8\pi\beta\kappa)^{-1}\ln(\kappa\Lambda_0^2)$ and
$C_3=1+\alpha_m$. As shown in Fig.~\ref{fig:fit}, this yields a nice fit
with $C_1-\alpha_0=0.0258\pm0.0003$ (with $\alpha_0$ the unknown
expansion at lowest pressure~\cite{Evans_pr}), $\beta\kappa=10.7\pm0.3$,
and $k_m/C_3=183\pm6\,{\rm erg}/{\rm cm}^2$.  The slope in the linear
low tension regime precisely determines $\beta\kappa$. The value of
$C_1$ is hard to interpret, since $\alpha_0$ and $\alpha_m$ are unknown.
An estimate of $k_m$ can be obtained by assuming $\alpha_m\simeq0.05$
(the range of the area expansion variation); this yields
$k_m\simeq192\,{\rm erg}/{\rm cm}^2$. In Ref.~\cite{Rawicz00}, the data
was also nicely fitted with Helfrich's formula~\cite{Helfrich84}, since
it equally leads to Eq.~(\ref{forfit}) in the tense regime, however with
different definitions for $C_1$ and $C_3$. These differences affect the
determination of $k_m$, since $C_3=1$ in Helfrich's theory.


\medskip{\em Conclusion\/}. -- Starting from a clearly defined picture at the
microscopic scale (an almost incompressible membrane built of a fixed number of
lipids, attached to a fixed frame, or enclosing a quasi-spherical volume), we
have derived an explicit Hamiltonian ${\cal H}_{\rm eff}[h]$ describing the
elasticity of a membrane, as gauged by its optically resolved shape~$h$.  This
effective description allows, in theory, to determine the response to an
external perturbation described by a Hamiltonian ${\cal H}_{\rm ext}[h]$, e.g.,
to a set of pulling micron-size beads. Our formulae for the effective tension
in the stretched states resemble those of Ref.~\cite{Helfrich84}, but our
formalism offers a clearer definition, a wider applicability, and basically a
justification of the concept of {\em effective tension}. This will allow us to
emphasize elsewhere the distinction between the effective membrane tension and
the mechanical frame tension $\Sigma=\beta^{-1}d(\ln{\cal Z})/dL^2$.

We have kept throughout the description at the Gaussian, quadratic level
(neglecting renormalization of the bending rigidities and associated
effects), however we believe that our results in their present form
offer a reasonably simple frame for the analysis of experiments,
including new micro-manipulation studies.

We thank D. Bartolo and C. Tordeux for illuminating discussions.
LP acknowledges the support of the CNRS.



\begin{references}
\bibitem{noteeq16}provided that $\kappa\Lambda_0^2$ is not much smaller
than $k_m$, which is indeed the case in Ref.~\protect\cite{Rawicz00}.
\end{references}
\end{document}